\documentstyle[11pt]{article}
\begin{document}
\newtheorem{lemma}{Lemma}[section]
\newtheorem{corollary}{Corollary}[section]
\newtheorem{theorem}{Theorem}[section]
\newcommand{\nc}{\newcommand}
\nc{\be}{\begin{displaymath}}
\nc{\ee}{\end{displaymath}}
\nc{\ble}{\begin{equation}}
\nc{\ele}{\end{equation}}
\nc{\bea}{\begin{eqnarray}}
\nc{\eea}{\end{eqnarray}}
\nc{\bela}{\begin{eqnarray*}}
\nc{\eela}{\end{eqnarray*}}
\nc{\todo}[1]{\par\noindent{\bf $\rightarrow$ #1}}
\def\twPsi{{\widetilde \Psi }}
\def\twC{{\widetilde C}}
\def\twcR{{\widetilde \cR}}
\def\lst{{l^*}}
\def\dst{{d^*}}
\def\bC{{\bf C}}
\def\bN{{\bf N}}
\def\bP{{\bf P}}
\def\bR{{\bf R}}
\def\bZ{{\bf Z}}
\def\cA{{\cal A}}
\def\cB{{\cal B}}
\def\cC{{\cal C}}
\def\cF{{\cal F}}
\def\cG{{\cal G}}
\def\cH{{\cal H}}
\def\cL{{\cal L}}
\def\cN{{\cal N}}
\def\cP{{\cal P}}
\def\cR{{\cal R}}
\def\cS{{\cal S}}
\def\cT{{\cal T}}
\def\cU{{\cal U}}
\def\cZ{{\cal Z}}
\def\om{{\overline m}}
\def\oU{{\overline U}}
\def\oX{{\overline X}}
\def\g{{\gamma}}
\def\L{{\Lambda}}
\def\rv{{\rm v}}
\def\eins{{\bf 1}}
\def\und#1{\underline{#1}}
\def\Gau#1#2{d\mu_{#1}({#2})}
\def\iGau#1#2{\int d\mu_{#1}({#2})}
\def\Gaug{d\mu_{\g}({\Phi })}
\def\Gaurv{d\mu_{\rv }({\Phi })}
\def\iGaug{\int d\mu_{\g}({\Phi })}
\def\GauN#1#2{d\mu_{#1}^{(N)}({#2})}
\def\iGauN#1#2{\int d\mu_{#1}^{(N)}({#2})}
\def\iGaugN{\int d\mu_{\g}^{(N)}({\Phi })}
\def\GaugN{d\mu_{\g}^{(N)}({\Phi })}
\def\GaurvN{d\mu_{\rv }^{(N)}({\Phi })}
\def\Ap{{A^{\prime }}}
\def\ap{{a^{\prime }}}
\def\Lp{{\Lambda^{\prime }}}
\def\Rp{{R^{\prime }}}
\def\Vp{{V^{\prime }}}
\def\Zp{{Z^{\prime }}}
\def\up{{u^{\prime }}}
\def\gp{{\g^{\prime }}}
\def\rhop{{\rho^{\prime }}}
\def\yp{{y^{\prime }}}
\def\inn{{\underline \in }}
\renewcommand{\thefootnote}{\arabic{footnote}}
%
\title{
  \begin{flushright} {\small MS--TPI--96--05} \end{flushright}
\vskip 2cm
A Convergence Proof\\
            for\\
            Linked Cluster Expansions}
\author{A. Pordt\thanks{email: pordt@poincare.uni-muenster.de}\\
Institut f\"ur Theoretische Physik I, Universit\"at M\"unster, \\
Wilhelm-Klemm-Str.\ 9, D-48149 M\"unster, Germany\\ [0.3cm]
}
\maketitle
\begin{abstract}
We prove that for a general $N$-component model
on a $d$-di\-men\-si\-o\-nal lattice $\bZ^d$
with pairwise nearest-neighbor
coupling and general local interaction obeying a stability
bound the linked cluster
expansion has a finite radius of convergence.
The proof uses Mayer Montroll equations for connected Green functions.
\end{abstract}

%

\newpage

\section{Introduction}

Linked cluster expansion is a useful tool to describe models in
the massive phase region \cite{Wor74}. Under the supposition that the first
singularity of this expansion is real and equal to the critical
point $\kappa_{crit}$ one can study the critical behaviour of the model
\cite{LueWei88} and \cite{Rei95}.
The high-order coefficients of the series expansion contains information
that enables one to compute critical exponents. These methods
were applied to study the ultraviolet resp. infrared
behaviour of Euclidean field theories \cite{LueWei88}.
We will here not prove that the radius of convergence is equal
to $\kappa_{crit}$. In this paper it will be shown that the
linked cluster expansion has a finite radius of convergence.

Let us consider a model with $N$-components on the lattice
$\Lambda = \bZ^d$ described by the partition function
\begin{equation} \label{e1}
Z(J,\kappa ) = \int \prod_{x\in \Lambda} d^N\Phi (x)\, \exp \{
   -S(\Phi , \kappa ) + \sum_{x\in \Lambda} J(x)\cdot \Phi (x)\} .
\end{equation}
We have used the scalar product
\be
J (x)\cdot \Phi (x):= \sum_{a=1}^N J_a(x) \Phi_a(x)
\ee
for $J,\Phi \in \cH (\Lambda )$, the Hilbert space of square summable
functions on $\Lambda .$
The action $S$ contains a local part $V$ and a non-local
pair interaction
\be
S(\Phi , \kappa ) = \sum_{x\in \Lambda} V(\Phi (x)) -
  \frac{1}{2} \sum_{x,y\in \Lambda} \sum_{a,b=1}^N
     \Phi_a (x) v_{ab}(x,y) \Phi_b(y).
\ee
The nearest-neighbor interaction is defined by
\begin{equation} \label{nnprop}
v_{ab}(x,y) = \left\{ \begin{array}{r@{\quad:\quad}l}
            2\kappa \delta_{a,b} & \mbox{$x,y$ nearest-neighbors} \\
            0  &  \mbox{otherwise.}
            \end{array}  \right.
\end{equation}
$\kappa $ is called the {\em hopping parameter}.
The partition function in eq. (\ref{e1}) is the generating function of
the Green functions. The generating function of the connected
Green functions reads
\be
W(J,\kappa ) = \ln Z(J,\kappa ).
\ee
The connected Green functions are
\be
G_{a_1\ldots a_n}^c(x_1,\ldots ,x_n) :=
  \frac{\partial^n}{\partial J_{a_1} (x_1) \cdots \partial J_{a_n}(x_n)}
   W(J,\kappa )\vert_{J=0},
\ee
for all $a_1, \ldots , a_n \in \{ 1,\ldots , N\} $ and $x_1,\ldots ,x_n \in
\Lambda .$ The linked cluster expansion is a Taylor expansion
around the local part of the action \cite{Wor74}. It uses successively
\be
\frac{\partial}{\partial v_{ab}(x,y)} W = \frac{1}{2}
(\frac{\partial^2W}{\partial J_a(x) \partial J_b(y)} +
  \frac{\partial W}{\partial J_a(x)} \frac{\partial W}{\partial J_b(y)} ).
\ee

We give the convergence proof only for the connected 2-point Green function.
Generalization to connected $n$-point Green functions are straightforward.

Furthermore, we will suppose translation invariance and $O(N)$-symmetry .
These assumptions are not essential for the proof and
generalization to more general cases are simply made.

We want to prove the following theorem
\begin{theorem} \label{Th1}
Let an $N$-component model in $d$ dimensions be defined by
the partition function in eq. (\ref{e1}). Suppose that there
exists a positive $c > 0$ and real $\delta $
such that for all $\Phi \in \bR^N$
\begin{equation} \label{Stab}
V(\Phi ) \ge c\Phi^2 - \delta.
\end{equation}
Then there exists a positive constant $\kappa_*$ such that
the 2-point Green function $G_{aa}^c(x,y)$ is an analytic function
in $K(\kappa_*) := \{ \kappa \in \bC |\, |\kappa | < \kappa_* \} $
and there exist for all $\kappa \in K(\kappa_*)$
positive $\alpha > 0$ and $m>0$ such that
for all $a\in \{ 1,\ldots ,N\}$, $x,y\in \Lambda $
\begin{equation} \label{Gbound}
|G_{aa}^c(x,y)| \le  e^\alpha \, q_V^{(2)}\delta_{xy} +
\alpha \, \exp \{ -m\Vert x -y\Vert \},
\end{equation}
where
\begin{equation} \label{q2def}
q_V^{(2)} := \frac{\int d^N\Phi \, \Phi^2_a \,
  \exp \{ -V(\Phi )\}}{\int d^N\Phi \, \exp \{ -V(\Phi )\}} .
\end{equation}
Furthermore, for $\kappa \rightarrow 0$,
\begin{equation} \label{Mass}
m(\kappa ) = O(\ln |\kappa |) > 0.
\end{equation}
\end{theorem}
Thus, using a well-known theorem for analytic functions,
the above theorem implies that for $m>0$ the series
in the definition of the 2-point susceptibility
\be
\chi_2 := \sum_{y\in \Lambda } G_{aa}^c(x,y)
\ee
exists and is an analytic function in $K(\kappa_*)$ if $\kappa_*$
is small enough.

In the remainder of this section we give an outline of the organization
of the proof.

The existence of the Green functions in the infinite volume $\Lambda $ is
not a priori assured. One can easily see that in a finite volume
(e.g. on a torus)
the derivatives of partition functions (not-normalized Green
functions) are well-defined quantities.
Furthermore,
by the dominated convergence theorem, they are entire
functions in $\kappa .$ We only have to show the existence
of the integral in the partition function. This is assured by
the stability bound (\ref{Stab}) for the local action.

Thus the first step in our convergence proof is to define the model
on finite subsets of the lattice and state the stability bound for
the local interaction $V.$ This restriction defines a polymer system
and polymer activities. Then we are in the position to express the
Green functions in terms of derivatives of polymer activities.

In a second step we formulate the thermodynamic limit of the Green
functions in terms of a series expansion
(Mayer Montroll equations) and show the convergence
under certain assumptions. This leads to a uniform bound of the
linked cluster expansion for small $\kappa .$ By a standard theorem
for series of analytic functions we can show that the linked
cluster expansion has a finite radius of convergence.
The Mayer Montroll equations avoid the problem of zeroes in the
partition function.
\section{Polymer system and stability}
In this section we derive the Mayer Montroll equations for the connected
2-point Green function (cp. \cite{MacPor89}) and show that the partition
functions for finite subsystems are well-defined if the local
interaction fulfills the stability bound.

{\em Polymers} are certain finite nonempty subsets of the lattice $\Lambda .$
The exact definition of the polymers is given below.
For a polymer $X\subset \Lambda $ define the partition function
\begin{equation} \label{e3}
Z(X|J,\kappa ) = \int \prod_{x\in X} d^N\Phi (x)\, \exp \{
   -S(X|\Phi , \kappa ) + \sum_{x\in X} J(x)\cdot \Phi (x)\} ,
\end{equation}
where the action is
\be
S(X|\Phi , \kappa ) = \sum_{x\in X} V(\Phi (x)) -
  \frac{1}{2} \sum_{x,y\in X} \sum_{a,b=1}^N
     \Phi_a (x) v_{ab}(x,y) \Phi_b(y).
\ee
The propagator kernel $v_{ab}(x,y)$ is zero unless $x,y$ are
nearest-neighbors and $a=b$ (cf. \ref{nnprop}).

By the dominated convergence theorem we may conclude that
$Z(X|J,\kappa )$ is analytic for all $\kappa \in \bC$
if the integral in eq. (\ref{e3}) exists.
The existence of the integral can be shown if the local interaction
satisfies the stability bound given by (\ref{Stab}).
This is easily seen by using the estimate
\bela
\lefteqn{|\frac{1}{2} \sum_{x,y\in X} \sum_{a,b=1}^N
\Phi_a (x) v_{ab}(x,y) \Phi_b(y)| =}
\nonumber\\  & &
|\sum_{<xy> :\, <xy> \in \cL (X)} (2\kappa ) \  \Phi(x) \cdot \Phi (y)| \le
    4d|\kappa |\sum_{x\in X} \Phi (x)^2.
\eela
$\cL (X)$ denotes the set of all links $<xy>$ ($x$ and $y$
are nearest-neighbors),
$x,y\in X$, in the lattice $\Lambda .$ The links have no directions,
i.e. $<xy> \equiv <yx>.$
\begin{lemma} \label{L1}
Suppose that there exists a positive $c>0$ and real $\delta $
such that the stability bound for the local interaction eq. (\ref{Stab})
is valid.
Then we have, for all $\kappa \in \bC $ and $\epsilon >0$ obeying
$4d|\kappa | + \epsilon < c, $
\be
S(\Phi ,\kappa ) \ge \epsilon \sum_x \Phi^2(x) - \delta
\ee
and the integral on the right hand side of eq. (\ref{e3}) is convergent.
\end{lemma}
Define polymer activities $A(Q|J,\kappa )$ for all polymers
$Q$ of $\Lambda $ such that the {\em polymer representation} of the
partition functions hold
\begin{equation} \label{e4}
Z(X|J,\kappa ) = \sum_{X=\sum Q} \prod_Q A(Q|J,\kappa )
\end{equation}
for all polymers $X$ of $\Lambda .$ The sum goes over all partitions of
$X$ into disjoint polymers. Eq. (\ref{e4}) defines the activities
uniquely. The reverse relation is
\begin{equation} \label{e5}
A(X|J,\kappa ) = \sum_{n\ge 1} (-1)^{n-1} (n-1)! \,
      \sum_{X=\sum_{i=1}^n Q_i} \prod_{i=1}^n Z(Q_i|J,\kappa ).
\end{equation}
Since $Z(X|J,\kappa )$ is analytic in $\kappa \in \bC $
under the supposition of lemma \ref{L1}
we see by eq. (\ref{e5}) that the
activities $A(X|J,\kappa )$ are also analytic functions.

The condition of nearest-neighbor couplings for the
non-local part of the action
implies that polymer activities are zero for non-connected polymers.
We say that a polymer $Q$ is {\em connected} if any $x,y\in Q$ can be
connected by links $l=<xy> \in \cL (X).$ Suppose that $X=X_1+X_2$
is the disjoint partition of $X$ into connected polymers $X_1$ and $X_2.$
Then the partition function factorizes
\be
Z(X|J,\kappa ) = Z(X_1|J,\kappa )\, Z(X_2|J,\kappa ).
\ee
This implies that $A(Q|J,\kappa )=0$ if $Q$ is not connected.
In the following we call finite nonempty subsets of $\Lambda $
{\em polymers} if they are connected.

A polymer $X\subset \Lambda $ with only one element, $|X| = 1,$ is
called {\em monomer}. The corresponding monomer activity is special. It is
equal to the monomer partition function
\begin{equation} \label{qdef}
A(\{ x\} |J,\kappa ) = Z(\{ x\} |J,\kappa ) =
\int d^N\Phi (x) \exp \{ -V(\Phi (x)) + J(x) \cdot \Phi (x) \} .
\end{equation}
For later convenience we will replace the partition functions
\be
Z(X |J,\kappa ) \longrightarrow
    \frac{Z(X |J,\kappa )}{\prod_{x\in X}Z(\{ x\} |J=0,\kappa )}.
\ee
Obviously, this is equivalent to the replacement of activities
\be
A(X |J,\kappa ) \longrightarrow
    \frac{A(X |J,\kappa )}{\prod_{x\in X}A(\{ x\} |J=0,\kappa )}.
\ee
This replacement only changes the normalization factor and will
not change the Green functions. After replacement we have
the normalization condition
\begin{equation} \label{e6}
A(\{ x\}  |J=0,\kappa ) = 1
\end{equation}
for all $x\in \Lambda .$

In the remainder of this section we want to express the connected
Green functions by polymer activities. We discuss this at the example
of the 2-point Green function. It is given by the (up to now formal)
thermodynamic limit
\bea \label{e7}
G_{aa}^c(x,y) &=& \lim_{X\nearrow \Lambda}
   \frac{\frac{\partial^2 Z(X|J,\kappa )}{\partial J_a(x) \partial J_a(y) }
  }{Z(X|J,\kappa )}\vert_{J=0}
\nonumber\\  &=&
\sum_{Q:\, Q\subseteq \Lambda}
\frac{\partial^2 A(Q |J,\kappa )}{\partial J_a(x) \partial J_a(y)}
  \vert_{J=0} \, \rho_\Lambda (Q|J=0,\kappa ),
\eea
where the {\em reduced correlation function} $\rho_X $ is defined by
\be
\rho_X (Q|J,\kappa ) := \frac{Z(X-Q|J,\kappa )}{Z(X|J,\kappa )},
\ee
for all polymers $Q,X,\  Q\subseteq X.$ Eq. (\ref{e7}) can be derived
by using the Kirkwood-Salsburg equation (cp. \cite{GruKun71})
\be
Z(X|J,\kappa ) = \sum_{Q:\, x\in Q\subseteq X} A(Q|J,\kappa )
                   Z(X-Q|J,\kappa )
\ee
for all $x\in X.$ This follows easily from the polymer representation
eq. (\ref{e4}).

Let us define {\em Mayer activities} $M$ by
\be
M(X|J,\kappa ) = -\delta_{1,|X|} + A(X|J,\kappa ).
\ee
The normalization condition eq. (\ref{e6}) is equivalent to
\be
M(\{ x\} | J=0,\, \kappa ) = 0
\ee
for all $x\in \Lambda .$

The Mayer Montroll equations express the reduced correlation functions
in terms of Mayer activities. For the formulation of these equations we
need some notations and definitions (see also \cite{MacPor89}).
Two polymers $P_1$ and $P_2$ are called compatible, $P_1 \sim P_2$, iff
they are disjoint, $P_1 \cap P_2 = \emptyset .$ A finite set
$\cP = \{ P_1, \ldots ,P_n\} $ consisting of polymers is called
admissible iff $P \sim P^{\prime}$ for all $P, P^{\prime} \in \cP ,$
$P \ne P^{\prime}.$ $K(X)$ denotes the set of all admissible $\cP $
which consists of polymers $P\subseteq X$
\be
K(X) := \{ \cP = \{ P_1, \ldots ,P_n\} |\,  \mbox{$\cP $ admissible,
    $P_i \subseteq X$} \} .
\ee
Let $\Pi (Y)$ be the set of all partitons of $Y$ into nonempty disjoint
subsets. Define $\Pi (\emptyset ) := \emptyset .$ Then we have
\be
K(X) = \bigcup_{Y:\, Y \subseteq X} \Pi (Y) .
\ee
Two admissible sets $\cP^{(1)} = \{ P_1^{(1)}, \ldots ,P_n^{(1)}\} $
and $\cP^{(2)} = \{ P_1^{(2)}, \ldots ,P_n^{(2)}\} $ are called compatible,
$\cP^{(1)} \sim \cP^{(2)}$ iff $P\sim P^{\prime}$ for all $P\in
\cP^{(1)}$, $P^{\prime} \in \cP^{(2)}.$

Denote by $Conn_X(\cP )$ the set which consists of admissible sets
$\cP^{\prime} \in K(X)$ which contain polymers that are incompatible
with at least one polymer $P$ in $\cP ,$ $P^{\prime} \not\sim \cP ,$
\be
Conn_X(\cP ) := \{ \cP^{\prime} \in K(X)|\, P^{\prime} \not\sim \cP \
   \forall P^{\prime} \in \cP^{\prime} \} .
\ee
We will use the notation $Conn(\cP ) \equiv Conn_\Lambda (\cP ).$
For a finite set $\cP = \{ P_1, P_2, \ldots \} $ define
\be
M^\cP := \prod_{P\in \cP } M(P|J,\kappa ).
\ee
The polymer representation eq. (\ref{e4}) implies
\be
Z(X|J,\kappa ) = \sum_{\cP :\, \cP \in K(X)} M^\cP .
\ee
We are ready to state the Mayer Montroll equations of the reduced
correlation functions, $\cP_0 := \{ \{ x\} |\, x\in X\} ,$
\begin{equation} \label{MM}
\rho_\Lambda (X|J,\kappa ) = \sum_{n\ge 0} \sum_{\cP_1,\ldots ,\cP_n \in
  K(X) \atop \emptyset \ne \cP_i \in Conn(\cP_{i-1}), i=1,\ldots ,n}
   (-M)^{\cP_1+\cdots +\cP_n}.
\end{equation}
For a proof see \cite{MacPor89}.
Insertion of eq. (\ref{MM}) into eq. (\ref{e7}) gives the connected
2-point function in terms of Mayer activities. It remains to
prove the thermodynamic limit of this expansion.
\section{Thermodynamic Limit}
In this section the thermodynamic limit of the Mayer Montroll
equation for the 2-point Green function is proven for small
complex $\kappa .$ This finishes the proof of the theorem \ref{Th1}.

Insertion of eq. (\ref{MM}) into eq. (\ref{e7}) yields
the Mayer Montroll equation for the 2-point Green function
\bea \label{e8}
G_{aa}^c(x,y) &=&
\sum_{\mbox{$Q:\, Q$ polymer}} \sum_{n\ge 0}
 \sum_{\cP_1,\ldots ,\cP_n \in K(Q) \atop \emptyset \ne \cP_i \in
  Conn(\cP_{i-1})} \frac{\partial^2 M(Q |J,\kappa ) }{\partial J_a(x)
      \partial J_a(y)}
\nonumber \\ & &
     (-M)^{\cP_1+\cdots +\cP_n}\vert_{J=0}.
\eea
The series starts with $\cP_0 = \{ \{ x\} |\, x\in Q\} .$
The individual terms in this series expansion of
$G_{aa}^c(x,y)$ are analytic
in $\kappa .$ We will show that this expansion is uniformly bounded
for all $|\kappa | \le \kappa_*$ by a convergent series expansion.
Then we may conclude by a standard theorem of complex functions that
$G_{aa}^c(x,y)$ is analytic in $\kappa $, $|\kappa | \le \kappa_*.$

For estimations we need the following tree graph formula for
the activities (cp. \cite{AbdRiv94}).
Let $T(X)$ be the set of all tree graphs with lines $(xy), \
x,y\in X$ and vertex set $X$. In the case of nearest-neighbor
interactions lines are links in $\cL (X).$ The tree graph formula
holds also for general pair interactions. For all polymers $X$ ,
$|X| \ge 2,$ we can write the polymer activity as a sum over
tree graphs $\tau$
\bea \label{Tree}
\lefteqn{A(X|J,\kappa ) = }
\nonumber\\ & &
\sum_{\tau :\, \tau \in T(X)} (2\kappa )^{|X|-1}
\int \prod_{x\in X} d^N\Phi (x) \,
\nonumber\\ & &
\exp \{ \sum_{x\in X}(-V(\Phi (x)) +
  J(x)\cdot \Phi (x) )\}\, [\prod_{(xy)\in \tau} \int_0^1 dt_{xy}]
\nonumber\\ & &
\prod_{<xy> \in \tau} (\Phi (x) \cdot \Phi (y)) \,
\exp \{ 2\kappa \sum_{<xy> \in \cL (X)} t_\tau^{\mbox{min}}(x,y) \,
  \Phi (x)\cdot \Phi (y) \} ,
\eea
where
\bela
\lefteqn{t_\tau^{\mbox{min}}(x,y) := }
\nonumber\\ & &
\mbox{min}\, \{ t_l|\, \mbox{ path
   connecting $x$ and $y$ and containing link $l\in \tau $} \} .
\eela
For a proof of this formula see \cite{AbdRiv94}

The following lemma shows under which conditions the Mayer Montroll
expansion for the connected 2-point Green function is convergent
and exponentially bounded. For convergence conditions of cluster
expansion and general polymer systems for the free energy
cf. \cite{KotPre86}.
\begin{lemma} \label{L2}
Suppose that there exists positive constants $\alpha ,\kappa_* , m
>0$ such that
\begin{equation} \label{e10}
\sum_{P:\, y\in P \atop |P| \ge 2} |M(P|J=0,\, \kappa )| \, \exp \{
  \alpha |P| \} \le \frac{\alpha }{2}
\end{equation}
and
\bea \label{e11}
\lefteqn{
\sum_{P:\, y\in P }
|\frac{\partial^2 M(P|J,\kappa )}{\partial J_a(x) \partial J_a(y)}
   \vert_{J=0} | \, \exp \{ \alpha |P| \} \le }
\nonumber \\ & &
   e^\alpha \, q_V^{(2)} \delta_{xy} +
   \frac{\alpha }{2}\, \exp \{ - m \Vert x-y\Vert \} ,
\eea
for all $x,y\in \Lambda $ and $|\kappa | \le \kappa_*,$
$a\in \{ 1,\ldots ,N\} ,$ where
\be
q_V^{(2)} := \frac{\int d^N\Phi \, \Phi^2 \,
  \exp \{ -V(\Phi )\}}{\int d^N\Phi \, \exp \{ -V(\Phi )\}} .
\ee
Then the series expansion eq. (\ref{e8}) is convergent and
\begin{equation} \label{e11a}
|G_{aa}^c(x,y)| \le  e^\alpha \,  q_V^{(2)}\delta_{xy} +
 \alpha \, \exp \{ - m \Vert x-y\Vert \} ,
\end{equation}
\end{lemma}
\par\noindent
{\em Proof :} We will show that
\begin{equation} \label{e12}
\sum_{\cP^{\prime}:\, \emptyset \ne \cP^{\prime} \in \,  Conn(\cP )}
  \exp \{ \alpha \Vert \cP^{\prime} \Vert \} \, |M|^{\cP^{\prime}} \le
      \frac{1}{2} \exp \{ \alpha \Vert \cP \Vert \}
\end{equation}
for all admissible $\cP = \{ P_1,P_2 ,\ldots \} $. We have used the
notation
\be
\Vert \cP \Vert := |P_1| + |P_2|  + \cdots .
\ee
The left hand side of eq. (\ref{e12}) is equal to
\begin{equation} \label{e13}
\sum_{n\ge 1} \sum_{P_1^{\prime} ,\ldots , P_n^{\prime} :\,
 P_a^{\prime} \not\sim \cP
\atop P_a^{\prime} \cap P_b^{\prime} =\emptyset ,\, a\ne b} \frac{1}{n!}
  \prod_{i=1}^n \exp \{ \alpha |P_i^{\prime}| \} \,
|M(P_i^{\prime}|J=0,\, \kappa )|.
\end{equation}
We use eq. (\ref{e10}) and the normalization condition eq. (\ref{e6})
to perform the sums over
$P_1^{\prime}, \ldots , P_n^{\prime}.$ Let us start with the sum over
$P_n^{\prime}.$ Using eq. (\ref{e10}) yields
a factor $(\Vert \cP \Vert -n+1)\frac{\alpha}{2}$ since
$P_n^{\prime}$ has to be incompatible with $\cP -
\{ P_1^{\prime} ,\ldots , P_{n-1}^{\prime} \} $ which consists of at most
$\Vert \cP \Vert -n+1$ elements. Then we sum over $P_{n-1}^{\prime}.$
This yields a factor $(\Vert \cP \Vert -n+2)\frac{\alpha}{2}.$ In the
same way we sum over $P_{n-2}^{\prime}, \ldots ,P_{1}^{\prime}.$ Thus
the term in (\ref{e13}) is bounded by
\be
\sum_{n \ge 1} {\Vert \cP \Vert \choose n} \, (\frac{\alpha}{2})^n \le
\frac{1}{2} [ (1+\alpha)^{\Vert \cP \Vert } -1 ] \le
\frac{1}{2} \, \exp \{ \alpha \Vert \cP \Vert \} .
\ee
This proves eq. (\ref{e12}). Using eq. (\ref{e12}) inductively, we obtain
\bea \label{e+}
|G_{aa}^c(x,y)| &\le &  e^\alpha \,  q_V^{(2)}\delta_{xy}
\nonumber \\ &+&
\sum_{P:\, y\in P } \sum_{n\ge 0} 2^{-n} \,
|\frac{\partial^2 M(P|J,\kappa )}{\partial J_a(x) \partial J_a(y)}
   \vert_{J=0} | \, \exp \{
  \alpha |P| \} .
\eea
The bounds (\ref{e11}) and (\ref{e+}) imply the assertion
(\ref{e11a}).  $\qquad  \Box  $

It remains to prove the supposition of lemma \ref{L2}.
\begin{lemma} \label{L3}
Suppose that the stability bound (\ref{Stab}) holds. For all $m>0$
there exists positive constants $\alpha >0$ and $\kappa_* >0$ such that
inequalities (\ref{e10}) and (\ref{e11}) are valid for all $|\kappa | \le
\kappa_* .$ Furthermore, for $\kappa \rightarrow 0,$
\be
m(\kappa ) = O(\ln |\kappa |) > 0.
\ee
\end{lemma}
\par\noindent
{\em Proof :} Using the stability bound (\ref{Stab}),
lemma \ref{L1} and the tree graph
formula (\ref{Tree}) for the polymer activities $A$ we obtain
for all polymers $X$, $|X| \ge 2,$
\begin{equation} \label{e14}
|M(X|J=0,\, \kappa)| \le \sum_{\tau :\, \tau \in T(X)} (2\kappa N)^{|X|-1}
   (e^\delta q_V)^{|X|} \epsilon^{-\frac{\sum_{x\in X}(d_\tau(x)+1)}{2}}
   \prod_{x\in X} (\frac{d_\tau(x)+1}{2})!,
\end{equation}
where
\be
d_\tau(x) := |\{ l\in \tau |\, \mbox{$l$ emerges from vertex $x$} \} |
\ee
is the number of lines in the tree graph $\tau$ which are connected
to vertex $x.$ We have used
\be
2\int_0^\infty dx\, \exp \{ -\epsilon x^2 \} x^n =
  \epsilon^{-\frac{n+1}{2}} (\frac{n+1}{2})!
\ee
and the definition
\be
q_V := \int d^N \Phi \, \exp \{ -V(\Phi ) \} .
\ee
Since the lines in the tree graph $\tau $ are links (=pairs of
nearest neighbors) the number of lines emerging from a vertex is
bounded by
\begin{equation} \label{ec1}
d_\tau (x) \le 2d .
\end{equation}
We will use
\begin{equation} \label{ec2}
\sum_{x\in X} d_\tau (x) =2(|X|-1).
\end{equation}
The number of random walks containing $l$ links starting from a site $x$
is equal to $(2d)^l.$ We can run through a tree graph $\tau \in T(X)$
by a random walk starting from a vertex $x\in X$ and visiting each
line $l$ two times and ending in vertex $x.$ Such a vertex contains
$2(|X|-1)$ lines. Thus the number of all tree graphs with vertex set
$X$ is bounded by
\begin{equation} \label{ec3}
|T(X)|\le (2d)^{2(|X|-1)}.
\end{equation}
By the same argument the number of polymers $X$ with $n$ elements
containing site $x\in \Lambda $ is bounded by
\begin{equation} \label{ec4}
|\{ X \subset \Lambda |\, \mbox{$X$ polymer, $|X|=n$, $x\in X$}\} |
                                                        \le (2d)^{2(n-1)}.
\end{equation}
Inequalitities (\ref{e14}), (\ref{ec1}), (\ref{ec3}) and
eq. (\ref{ec2}) imply, for $|X| \ge 2$,
\bea \label{e15}
|M(X|J=0,\, \kappa )| &\le & (2(2d)^2 \kappa N)^{|X|-1}
(e^\delta q_V^{-1})^{|X|}
  \epsilon^{-\frac{3|X|-2}{2}} (d+\frac{1}{2})!^{|X|}
\nonumber\\  &=&
\frac{e^\delta (d+\frac{1}{2})!}{q_V \epsilon^{1/2}}
(\frac{2 (2d)^2\kappa e^\delta
N (d+\frac{1}{2})!}{q_V \epsilon^{3/2}})^{|X|-1}.
\eea
Similarly, we derive, for $|X| \ge 2,$
\begin{equation} \label{e16}
|\frac{\partial^2 M(X|J,\kappa )}{\partial J_a(x) \partial
J_a(y)} \vert_{J=0} |
 \le
  \frac{e^\delta (d+\frac{1}{2})!(1+\frac{1}{2d})^2}{q_V \epsilon^{3/2}}
     (\frac{2 (2d)^2\kappa e^\delta
     N (d+\frac{1}{2})!}{q_V \epsilon^{3/2}})^{|X|-1}.
\end{equation}
For the monomer case $X = \{ x \} $, $|X|=1$, we have,
using definition (\ref{q2def})
\begin{equation} \label{e17}
|\frac{\partial^2 M(X|J,\kappa )}{\partial J_a(x) \partial
J_a(y)} \vert_{J=0} |
= q_V^{(2)} \delta_{xy}
\end{equation}
for all $y\in \Lambda $, $x\in X.$
Eqs. (\ref{e15}) and (\ref{e16}) imply, for $|\kappa | \le \kappa_*$,
$\kappa_*$ small enough,
\begin{equation} \label{e18}
\sum_{P:\, x\in P} |M(P|J=0,\, \kappa )| \exp \{ \alpha |P| \} \le
  \frac{2 (d+\frac{1}{2})! e^{\alpha + \delta}}{q_V \epsilon^{1/2}} u,
\end{equation}
and
\begin{equation} \label{e19}
\sum_{P:\, x\in P\atop |P| \ge 2}
|\frac{\partial^2 M(P|J,\kappa )}{\partial J_a(x)
\partial J_a(y)} \vert_{J=0} |
\le  \frac{2 (1+\frac{1}{2d})^2(d+\frac{1}{2})! e^{\alpha
  + \delta}}{q_V \epsilon^{3/2}} u\,
\exp \{ -m\Vert x-y\Vert \} ,
\end{equation}
where
\be
u:= \frac{2(2d)^4 \kappa_* N (d+\frac{1}{2})!
  e^{\alpha + \delta}}{q_V \epsilon^{3/2} e^{-m}} \le
  \frac{1}{2} .
\ee
The factor in eq. (\ref{e19}) $\exp \{ -m\Vert x-y\Vert \} $ comes from
the fact that the terms on the left hand side of eq. (\ref{e19})
are zero unless $x,y\in P.$

For $\kappa_*$ small enough there exists a positive constant $\alpha > 0$
such that
\begin{equation} \label{e20}
\max (\frac{2 (d+\frac{1}{2})! e^{\alpha + \delta}}{q_V \epsilon^{1/2}} u,
\frac{2 (1+\frac{1}{2d})^2(d+\frac{1}{2})!
e^{\alpha + \delta}}{q_V \epsilon^{3/2}} u) \le
\frac{\alpha}{2}.
\end{equation}
Inequalities (\ref{e18}) and (\ref{e20}) proves the bound (\ref{e10}).
Inequalities (\ref{e17}), (\ref{e19}) and (\ref{e20}) imply
the bound (\ref{e11}).
Furthermore, we can take $m$ very large if $\kappa_* e^m$ remains small.
This proves that $m$ is of order $\ln (|\kappa | )$
for small $|\kappa |,$ eq. (\ref{Mass}). $\qquad \Box $
\section{Summary}
The partition functions restricted to finite connected subsets of
the lattice are analytic functions in the hopping parameter $\kappa .$
By the polymer representation the Mayer activities are also analytic.
The Mayer Montroll equations express the connected Green functions
by Mayer activities. If the local part of the action obeys a stability
condition it turns out that the linked cluster expansion of the
Green functions can be uniformly bounded using these Mayer
Montroll equations. The result is analyticity of the connected
Green functions and susceptibilities for $\kappa $ in a small
neighborhood of zero.

The proof presented here is restricted to nearest-neighbor interaction.
In a forthcoming paper \cite{PorRei96} the convergence proof will
be generalized to linked cluster expansions beyond nearest neighbor
interactions.

\section*{Acknowledgement}
I would like to thank T. Reisz for encouraging me to write this proof
and the Deutsche For\-schungs\-ge\-mein\-schaft for financial support.
\end{document}